\documentclass[prd, twocolumn, showpacs, superscriptaddress, floatfix, nofootinbib, preprintnumbers, amsmath, amssymb]{revtex4}

\usepackage{latexsym}
\usepackage{graphicx}

\usepackage{color}

\begin{document}

\title{Inflation in Entropic Cosmology: Primordial Perturbations and non-Gaussianities}

\author{Yi-Fu Cai}
\email[Email:]{ycai21@asu.edu}
\affiliation{ Department of Physics, Arizona State University, Tempe, AZ 85287, USA}

\author{Emmanuel N. Saridakis}
\email[Email:]{msaridak@phys.uoa.gr}
\affiliation{ College of Mathematics and Physics, Chongqing University of Posts and Telecommunications,\\ Chongqing 400065,
China}

\begin{abstract}
We investigate thermal inflation in double-screen entropic cosmology. We find that its realization is general, resulting from the system evolution from non-equilibrium to equilibrium. Furthermore, going beyond the background evolution, we study the primordial curvature perturbations arising from the universe interior, as well as from the thermal fluctuations generated on the holographic screens. We show that the power spectrum is nearly scale-invariant with a red tilt, while the tensor-to-scalar ratio is in agreement with observations. Finally, we examine the non-Gaussianities of primordial curvature perturbations, and we find that a sizable value of the non-linearity parameter is possible due to holographic statistics on the outer screen.
\end{abstract}


\maketitle



\section{Introduction}

As early as the study of black hole physics \cite{Bekenstein:1973ur, Hawking:1974sw}, the holographic thermodynamics was discovered to be related to the quantization of
Einstein gravity as a nonperturbative quantum feature. In particular, the holographic principle was conjectured as a significant property of quantum gravity, stating that physics of a volume of space is encoded on its boundary, such as a gravitational horizon \cite{'tHooft:1993gx}. This principle was also applied in cosmology \cite{Gibbons:1977mu, Bousso:2002ju} and it was studied in detail in string theoretical background \cite{Maldacena:1997re}.

Based on these, an extended holographic picture was suggested by Verlinde \cite{Verlinde:2010hp} in which Einstein gravity is no longer a fundamental theory, but it emerges from a statistic effect of a holographic screen, while a similar scenario was discussed by Padmanabhan \cite{Padmanabhan:2009vy}. The cosmological
application was extensively studied in the literature, for example see Refs. \cite{Cai:2010hk, Li:2010cj, Zhang:2010hi} and references thereafter. However, this
theory involves the controversial issue of whether the uniqueness of gravity is preserved in such an emergent scenario. Therefore, a more explicit formulation of entropic gravity theory was suggested \cite{Easson:2010av, Easson:2010xf}, in which Einstein gravity is still a fundamental theory but with a boundary term being introduced. Such a boundary term provides a holographic statistics and thus it leads to an entropic force in bulk physics. This model was soon applied to realize the current acceleration \cite{Easson:2010av} and inflation \cite{Easson:2010xf} at early times, but it has also led to some criticism from the point of view of observations \cite{Danielsson:2010uy}.

On the other hand, inflation has been widely considered as a remarkably successful theory in describing the very early universe\cite{Guth:1980zm}. In this paradigm, the primordial curvature perturbation caused by the quantum fluctuations of the inflaton field was found to be nearly scale-invariant and thus it is able to form the Large Scale Structure (LSS) of our universe \cite{Mukhanov:1990me}. Currently many observations, particularly the angular spectrum of the Cosmic Microwave Background (CMB) anisotropies \cite{Komatsu:2010fb} and the power spectrum of density fluctuations observed for the LSS \cite{LSS}, strongly support the compatibility of inflationary cosmology for describing the early universe.

Recently, an explicit scenario of realizing the inflationary period in entropic cosmology was proposed in \cite{Cai:2010zw}, composed by two holographic screens. In particular, it was found that inflationary solutions can be achieved even in a radiation dominated universe, provided the two screens are not at thermal equilibrium. Such realizations of ``thermal inflation'' have become an interesting issue in recent studies of entropic inflationary cosmology \footnote{See also \cite{Wang:2010jm, Li:2010bc} for relevant discussion in Verlinde's framework.}.

In the present work we are interested in investigating thermal inflation in double-screen entropic cosmology, both at the background as well as at the perturbations level. In particular, after showing the generality of inflationary solutions at high energy scales, we study the primordial curvature perturbations. As
we will see, the main contribution arises from the holographic fluctuations generated on the outer screen, while the usual thermal fluctuations of the universe content is subdominant, and the resulting a power spectrum is nearly scale-invariant with a red tilt. Additionally, by examining the non-Gaussianities for holographic initial conditions, we find that a sizable non-linearity parameter could be obtained.

The outline of this paper is as follows. In Section \ref{dbscrentr} we briefly review the scenario of entropic cosmology with two holographic screens, focusing on the background evolution, and in Section \ref{therminfl} we examine the realization of thermal inflation in this model. In Section \ref{primperturb} we perform an analysis of the cosmological perturbations generated during thermal inflation, which are mainly of holographic origin. Then, in Section \ref{non-gauss} we estimate the non-Gaussianities that arise in the examined scenario. Finally, Section \ref{Conclusion} is devoted to the summary of the obtained results.

\section{A double-screen model of Entropic cosmology}
\label{dbscrentr}

In this work we are interested in investigating thermal inflation
in a scenario of entropic cosmology involving two holographic
screens\cite{Cai:2010zw}. However, let us first remind the basic
features of standard, one-screen, entropic cosmology, which is
also called {\it EFS} scenario \cite{Easson:2010av,
Easson:2010xf}.

\subsection{One-screen entropic cosmology}

In usual entropic cosmology one incorporates a gravitational
system including matter fields and surface terms of the form of
\begin{eqnarray}
 {\cal I} = \int_{M_b}\bigg(\frac{R}{16\pi G}+{\cal L}_m\bigg) +\oint_{\partial {M_b}}{\cal L}_b~,
\end{eqnarray}
where $R$ is the Ricci scalar of the whole spacetime, ${\cal L}_m$
is the Lagrangian of matter fields living in the bulk, and ${\cal
L}_b$ is the corresponding Lagrangian describing the physics of
the boundary. Clues from string theory and AdS/CFT indicate that
the boundary terms should include the extrinsic curvature of the
boundary and holographic dual gauge theories. Finally, throughout
the paper we use the convention $c=k_B=\hbar=1$ and
$M_p=1/\sqrt{G}$.

By varying the action with respect to the metric, one obtains the
Einstein field equation as follows,
\begin{eqnarray}
 R^{\mu\nu}-\frac{1}{2}Rg^{\mu\nu}
  = 8\pi GT_m^{\mu\nu} + J_b^{\mu\nu}~,
\end{eqnarray}
in which the last term $J_b$ is a current describing the exchange
of energy and momentum between the bulk and the boundary. This
term is determined by the holographic description of boundary
physics, and so is a nonlocal effect which corresponds to an
entropic force in the universe.

Assuming that the boundary physics can be described by
thermodynamics satisfying a holographic distribution, the number
of degrees of freedom on this holographic screen is proportional
to its area, that is $N\propto A$. Thus, the classical holographic
entropy on this screen is given by
\begin{eqnarray}\label{Sb}
 S_b=\frac{A}{4G}=\frac{\pi}{G} r_b^2~,
\end{eqnarray}
where $r_b$ is the radius location of the boundary surface.
Therefore, variation of energy with respect to the radius will
provide us the entropic force  \cite{Verlinde:2010hp}:
\begin{eqnarray}\label{Fe}
 F_{e}=-\left(\frac{dE}{dr}\right)_b=-\left(T\frac{dS}{dr}\right)_b=-\frac{2\pi}{G} T_br_b~,
\end{eqnarray}
in which $T_b$ is the temperature of the boundary of the system.
Finally, due to the Unruh effect (when a test particle with mass
$m$ is located nearby the holographic screen the variation of the
entropy on this screen with respect to the radius takes the form
of $\frac{dS}{dr}=-2\pi m$) the above force yields an entropic
acceleration $a_e$ of the form  \cite{Unruh:1976db}
\begin{eqnarray}\label{ae}
 a_e\equiv\frac{F_e}{m}=2\pi T_b~.
\end{eqnarray}
Note that the corresponding entropic pressure is negative
$P_e=F_e/{A_b}=-{T_b}/{2Gr_b}$, and so it is expected to realize
an accelerating process of the universe.

Let us apply the above results into a homogeneous and isotropic
flat Friedmann-Robertson-Walker (FRW) universe described by the
metric
\begin{eqnarray}
 ds^2=dt^2-a(t)^2dx^idx^i~.
\end{eqnarray}
In usual, one-screen entropic cosmology, the boundary $r_b$, that
is the location of the holographic screen, is assumed to be near
the Hubble horizon $r_H=H^{-1}$, where $H\equiv \dot a/a$ is the
Hubble parameter of the universe. This non-complete coincidence is
quantified by the parameter $\beta$ \cite{Cai:2010zw}, that is we
write
\begin{eqnarray}
\label{hubblerad}
 r_b=(\beta H)^{-1},
\end{eqnarray}
 while the boundary temperature is
\begin{eqnarray}
\label{hubbletemp}
 T_b=\frac{\beta H}{2\pi}.
\end{eqnarray}
 Thus, substituting everything in the field
equations, we obtain the modified Friedmann acceleration equation
\begin{eqnarray}
\frac{\ddot a}{a}=-\frac{4\pi G}{3}(\rho+3p)+\beta^2H^2,
\end{eqnarray}
where $\rho$ and $p$ are respectively the total energy density and
pressure of the content of the universe. In this expression, the
last term accounts for the cosmological acceleration due to the
entropic force.

A final addition must be made, concerning the precise form of the
horizon entropy. In particular, quantum gravitational and string
theoretical considerations, taking into account higher order
quantum corrections \cite{Strominger:1996sh} and the holographic
renormalization group flow \cite{Alvarez:1998wr}, yield an
improved relation for the entropy with leading order correction
as:
\begin{eqnarray}
\label{Squantcorr}
 S=\frac{1}{4G}\left(A+gG\ln\frac{A}{G}+...\right),
\end{eqnarray}
where the coefficient $g$ is determined by the specific
environment and it is left as a free parameter. Thus, the
Friedmann acceleration equation arising from this improved
entropic relation reads  \cite{Easson:2010xf}
\begin{eqnarray}\label{HFRW1impr}
 \frac{\ddot a}{a} = -\frac{4\pi G}{3}(\rho+3p) + \beta^2H^2+ \frac{gG\beta^4H^4}{4\pi}+...~.
\end{eqnarray}

Although the aforementioned scenario is qualitatively very
interesting, the above modified Friedmann equation (with $\beta^2$
of the order of ${\cal{O}}(1)$) cannot quantitatively describe the
radiation and matter epochs. One interesting way out is the
additional consideration of a second holographic screen.

\subsection{Double-screen entropic cosmology}

Since one-screen considerations exhibit difficulties in
quantitatively describing the thermal history of the universe, a
double-screen extension was introduced in \cite{Cai:2010zw}. Since
the Hubble horizon (or a surface near it) is the natural choice
for the outer boundary of the universe, one introduces an
additional ``inner'' boundary, which is just the Schwarzschild
horizon of the whole universe. The corresponding Schwarzschild
radius $r_S$ is given by
\begin{eqnarray}
 r_S=2GM_{tot}=2G\int_{M_b}\rho dV=\frac{8\pi G\rho}{3\beta^3H^3},
\end{eqnarray}
where we have used that the volume of the universe is $V=4\pi
r_b^3/3$. Its corresponding temperature is given by
\begin{eqnarray}
 T_S=\frac{1}{8\pi G M_{tot}}=\frac{3\beta^3H^3}{32\pi^2G\rho},
\end{eqnarray}
and therefore its induced acceleration (with the simple entropy
form) will be
\begin{eqnarray}\label{aes}
 a_e=2\pi T_S,
\end{eqnarray}
but with direction towards the inner screen, that is opposite to
the outer one.

In summary, in double-screen entropic cosmology, the induced
acceleration is
\begin{eqnarray}
 a_e=2\pi (T_b-T_S)
 =\beta H\bigg(1-\frac{3\beta^2H^2}{16\pi G\rho}\bigg),
\end{eqnarray}
that is it incorporates a competition of entropic effects from the
outer and the inner screens. Consequently, the modified Friedmann
acceleration equation in this scenario writes as \cite{Cai:2010zw}
\begin{eqnarray}\label{HFRdbscr}
 \frac{\ddot a}{a} = -\frac{4\pi G}{3}(\rho+3p) + \beta^2H^2\bigg(1-\frac{3\beta^2H^2}{16\pi
 G\rho}\bigg).
\end{eqnarray}
Finally, if instead of the simple entropy form we use the quantum
corrected one (\ref{Squantcorr}), the modified Friedmann equation
in double-screen cosmology becomes
\begin{eqnarray}\label{HFRdbscrimpr}
 \frac{\ddot a}{a} = -\frac{4\pi G}{3}(\rho+3p) + f(\rho, H)~,
\end{eqnarray}
with the form of surface function being
\begin{eqnarray}
 && f(\rho, H)
  = \beta^2H^2\bigg(1-\frac{3\beta^2H^2}{16\pi G\rho}\bigg)\ \ \
  \ \ \ \ \ \ \ \ \ \ \  \ \ \ \ \ \   \ \ \ \ \ \  \ \ \
  \nonumber\\
  &&\ \ \  \ \  \ \  \
  \ +
\frac{g_HG\beta^4H^4}{4\pi}\left(1-\frac{27g_S\beta^6H^6}{1024g_H\pi^3G^3\rho^3}\right)+...~,
\end{eqnarray}
where $g_H$ and $g_S$ are the corresponding correction coefficient
for each boundary.

Eq. (\ref{HFRdbscrimpr}) determines the cosmological evolution in
double-screen cosmology. If the two holographic screens are in
thermal equilibrium with $T_b=T_S$ and choosing the coefficient
$\beta=\sqrt{2}$, one can recover the exact form of the
traditional Friedmann equation. However, in general, Eq.
(\ref{HFRdbscrimpr}) describes the evolution of the universe
towards such an equilibrium. The cosmological system will close,
as usual, by the consideration of the evolution equation of the
total energy density $\rho$. In the case at hand, in which one may
have flow through the boundaries, the corresponding equation is
modified as \cite{Cai:2010zw}
\begin{eqnarray}\label{HFRDBSCRIMPR2}
 \dot\rho+3H(\rho+p)=\Gamma~,
\end{eqnarray}
with the effective coupling term $\Gamma$ being
\begin{eqnarray}
 \Gamma = \frac{27\beta^6H^6}{1024\pi^3G^3\rho^3}\dot\rho
  +\frac{3\beta^2H\dot{H}}{4\pi G}
  \bigg(1-\frac{27\beta^4H^4}{256\pi^2G^2\rho^2}\bigg)~,
\end{eqnarray}
at classical level. Again, when $T_b=T_S$ and $\beta=\sqrt{2}$,
the coupling $\Gamma$ vanishes and (\ref{HFRDBSCRIMPR2}) takes its
standard form.

We close this subsection by mentioning the following. At early
cosmological times the aforementioned scenario holds as it is.
However, for completeness we mention that at late times, in order
to describe the dark-energy epoch and universe acceleration, one
has to take into account the evaporation of the inner,
Schwarzschild screen \cite{Cai:2010zw}. Since in the present work
we are interested in very early times, that is in inflationary
epoch, we will not make such a consideration in the following.

\section{Thermal inflation at early universe}
\label{therminfl}

In the previous section we analyzed the basic features of
double-screen entropic cosmology. Here we focus on the early-time
universe evolution, and in particular we examine the inflation
realization. Let us first show why such a realization is easily
obtained in the model at hand.

In such early-time epochs, the universe is radiation dominated,
and thus in the following we assume that the equation of state of
the total universe content is $p=\rho/3$. Solving the equations of
motion (\ref{HFRdbscrimpr}) and (\ref{HFRDBSCRIMPR2}) up to
leading order, considering the first order quantum correction to
the entropy, one can obtain the following approximate solution for
the Hubble parameter at early times  \cite{Cai:2010zw}
\begin{eqnarray}\label{HFRW}
 H^2 = \frac{8\pi G}{3}\bigg[\rho+\frac{8g}{69}G^2\rho^2+...\bigg],
\end{eqnarray}
where we have introduced the coefficient $g=g_H-4g_S$. Therefore,
the standard Friedmann equation can be achieved when $g=0$ at
early times. An interesting property of this scenario is that when
$g>0$, the Hubble parameter is proportional to the energy density
at high energy scales. In this case the $\rho^2$ term could make
the early time inflation much easier to be realized, providing an
implement of holographic inflation.

Let us now investigate the inflation realization in more detail.
In the case of $g>0$, at sufficiently early times the
$\rho^2$-term in (\ref{HFRW}) will always dominate, and thus the
universe will exhibit the inflationary epoch, which, since it is
radiation dominated, we call thermal inflation. This is a
difference from other examinations of inflation in one-screen
entropic cosmology, in which the neglecting of thermal effects
makes inflation difficult \cite{Easson:2010xf} or impossible
\cite{Li:2010bc}. As time passes and the universe grows, $\rho$
will be decreasing, and when it reaches the critical value of
$\rho_C\simeq 69/(8gG^2)$ the $\rho$-term will dominate,
triggering the end of inflation. Solving the equations of motion
one finds that the energy density of the universe evolves as
\cite{Cai:2010zw}
\begin{eqnarray}\label{inflation1}
 \rho\simeq\sqrt{\rho_C^2-\frac{512\pi^3t}{27g^{5/2}G^{9/2}}}.
\end{eqnarray}
In this relation the initial Big Bang time is set to be negative
infinity, the observable entropic thermal inflation starts at a
time $-\infty <t_{i}\ll0$ (surely $-g^{3/2}G^{1/2} \lesssim t_{i}$
since only after that time the Hubble parameter becomes smaller
than the Planck scale), while it ends at $t_C=0$, after which
$\rho$ becomes smaller than $\rho_C$ and standard
post-inflationary cosmology begins.

Proceeding forward one finds that at early times ($t\ll0$) the
Hubble parameter behaves like
\begin{eqnarray}
\label{Htapprox}
 H(t) \simeq 24.25\times\frac{(-t)^{1/2}}{(gG)^{3/4}},
\end{eqnarray}
and thus the slow-roll parameter $\epsilon$ reads:
\begin{eqnarray}
\label{epsilondef}
 \epsilon \equiv -\frac{\dot H}{H^2}
 \simeq 2.06\times10^{-2}\frac{(gG)^{3/4}}{(-t)^{3/2}},
\end{eqnarray}
which is indeed much less than unity when $t\ll-\sqrt{gG}$. Thus,
one can make an estimation for the efolding number ${\cal N}$, for
the observable inflationary stage starting at $t_{i}$ and ending
at $t_C=0$, as
\begin{eqnarray}
\label{efolds}
 {\cal N} \equiv \int^{t_{C}}_{t_{i}} H(t)dt
   \simeq 16.17\times \frac{(-t_{i})^{3/2}}{(gG)^{3/4}}~.
\end{eqnarray}
We mention that this is an approximate result, since the relation
(\ref{Htapprox}) does not hold up to $t=t_C=0$. Finally, note that
in the above case $g\sim{\cal{O}}(10^{16})$ \cite{Cai:2010zw}, as
it is implied by the requirement that the inner holographic screen
evaporates within the age of our universe so that one can obtain
the late-time acceleration. The relevant observational constraints
on this parameter will be studied in detail in near future.

\section{Primordial perturbations in entropic cosmology}
\label{primperturb}

In the previous section we investigated the realization of thermal
inflation in a double-screen entropic cosmology. The whole
analysis remained at the background level, since it is the one
that determines the basic features of the cosmological evolution.
In the present section we extend our analysis at the perturbation
level, since such an examination reveals important details of a
cosmological scenario. More importantly, especially for the case
of inflation, the perturbation analysis can be straightforwardly
confronted by observations, leading to strong constraints or
ruling out a specific inflationary model.

The standard mechanism of generating primordial perturbations is
to require that the initial cosmological fluctuations emerge
inside the Hubble radius, and subsequently they are transformed
into classical perturbations, through decoherence, after exiting
the Hubble radius. It is usually suggested that these initial
fluctuations are generated as quantum vacuum perturbations.
However, in the scenario of the present work, the universe is
always filled with radiation, even at very early times. As a
consequence, and as predicted by thermal field theory
\cite{Kraemmer:2003gd}, the thermal fluctuations dominate the
quantum ones, and thus their investigation is sufficient. Now, in
our case the thermal fluctuations have two origins, one is the
thermal particle fluctuation inside the bulk-universe, and the
other is the holographic fluctuation on the two boundary screens.
In the following we assume that the correlation between thermal
particle fluctuation and holographic perturbation is negligible,
and thus we calculate the contribution of each component
independently.

Thermal fluctuations as the origin of the structure in the
universe were considered in the context of an expanding universe,
but it was concluded that a scale-invariant spectrum of
cosmological perturbations could not be created from a usual
thermal origin  \cite{Magueijo:2002pg}. However, motivated by
string gas cosmology  \cite{Brandenberger:1988aj}, people have noticed that thermal
fluctuations satisfying a specific holographic statistical
distribution are able to provide a scale-invariant spectrum in
certain backgrounds, namely in a Hagedorn phase
\cite{Nayeri:2005ck}, in an eternally expanding universe
\cite{Magueijo:2006fu} and in bouncing cosmology
\cite{Cai:2009rd}. As we will show, this is not the case in the
scenario at hand, that is we can obtain a scale-invariant spectrum
without the need of specific considerations.

\subsection{The formalism}

We are interested in studying primordial curvature perturbations
originating both from the fluctuations of normal radiation and of
boundary matter on the two holographic screens. We start by
considering the perturbed flat FRW metric in longitudinal gauge,
which takes the usual form
\begin{eqnarray}
 ds^2=a(\tau)^2[(1+2\Phi)d\tau^2-(1-2\Phi)dx^idx^i],
\end{eqnarray}
where $\tau$ is the conformal time, and $\Phi(\tau, x^i)$
represents the metric fluctuation. Following the formula developed
in \cite{Cai:2009rd}, the key constraint equation relating matter
and metric fluctuations is given by the time component of the
perturbed Einstein equations, namely from
\begin{eqnarray}
 -3{\cal H}({\cal H}\Phi+\Phi')+\nabla^2\Phi=4\pi Ga^2\delta\rho,
\end{eqnarray}
where ${\cal H}=aH$ is the conformal Hubble parameter, the prime
denotes the derivative with respect to conformal time, and
$\delta\rho$ is the fluctuation of energy density which contains
thermal particle modes and holographic boundary ones. Finally, as
usual, one transforms into Fourier space, and uses the
corresponding modes as the relevant variables.

In summary, for a cosmological system filled with general thermal
matter, the thermally originated power spectrum $\Phi_k$ can be
expressed as  \cite{Cai:2009rd}
\begin{eqnarray}\label{PPhi}
 P_{\Phi}(k)\equiv\frac{k^3}{2\pi^2}\langle\Phi_k^2\rangle
  =\frac{8G^2\langle\delta\rho^2\rangle}{H^4}|_{t_*(k)}~,
\end{eqnarray}
up to a constant of order ${\cal{O}}(1)$, where $t_*(k)$ denotes
the moment of Hubble crossing for the specific mode. In this
expression, $\langle\delta\rho^2\rangle$ is the correlation
function of density fluctuations in position space, within a
sphere of radius $R(k)$, where $R(k)$ is the physical correlation
length corresponding to the co-moving momentum scale $k$.
Moreover, in a thermal system $\langle\delta\rho^2\rangle$ is
given by
\begin{eqnarray}
\label{deltarho2}
 \langle\delta\rho^2\rangle=C_V\frac{T^2}{R^6},
\end{eqnarray}
where $C_V\equiv\partial\langle{E}\rangle/\partial{T}$ is the heat
capacity of radiation matter. We mention that in our scenario
there exist two kinds of thermal matter, one being the normal
radiation constituted by a gas of relativistic point particles,
and the other being the boundary matter on the two holographic
screens. In the following two subsections we study the curvature
perturbations arisen from these two sources separately.

\subsection{Fluctuations from normal radiation}

In this subsection we consider primordial curvature perturbations
induced by the radiation sector that fills the universe during
thermal inflation. As it is known from thermodynamics, the
radiation energy density as a function of the temperature is given
by
\begin{eqnarray}
 \rho_r \sim T_r^4,
\end{eqnarray}
while the heat capacity of normal radiation reads
 \cite{Cai:2009rd}
\begin{eqnarray}\label{CVr}
 C_V^r=g_rR_r^3T_r^3,
\end{eqnarray}
where the subscript r stands for ``radiation'' and $R_r$ is the
radiation correlation length, given as usual from  $R_r=c_s/H$,
with $c_s$ the sound speed. Additionally, the coefficient $g_r$
characterizes the species of the relativistic point particles of
the radiation sector, and it usually takes a value of the order
${\cal{O}}(1)$.

Inserting (\ref{CVr}) in (\ref{deltarho2}) and then in
(\ref{PPhi}), with all quantities considered with a subscript $r$,
one obtains the expression of the power spectrum for metric
perturbations seeded by normal radiation, namely
\begin{eqnarray}
\label{radspect1}
 P_{\Phi}^r=\frac{g_r\beta^5}{4\pi^5c_s^3}G^2H^4.
\end{eqnarray}
Note that in the extraction of this relation we have assumed that
the background temperature of the universe is $T_r=\beta{H}/2\pi$
near thermal equilibrium, that is it coincides with the
temperature of the outer holographic screen given by
(\ref{hubbletemp}). Specifically, in a realistic model with
$c_s=1/\sqrt{3}$ and $\beta=\sqrt{2}$, (\ref{radspect1}) yields,
\begin{eqnarray}
\label{radspect2}
 P_{\Phi}^r=\frac{3\sqrt{6}g_r}{\pi^5}G^2H^4~.
\end{eqnarray}
Therefore, as can be clearly seen from (\ref{radspect1}) or
(\ref{radspect2}), the spectrum of curvature perturbation from
radiation fluctuations is scale invariant during thermal
inflation.

\subsection{Fluctuations from outer holographic screen}

In this subsection we consider primordial curvature perturbations
induced by holographic fluctuations on the outer screen. In
entropic cosmology the boundary terms satisfy a holographic
statistics, which states that the fundamental degrees of freedom
are bounded by their surface areas. According to the equipartition
principle, one can acquire the total energy of the outer screen as
$\langle{E}\rangle\sim r_bT_b/G$, with the boundary location $r_b$
and temperature $T_b$ given by (\ref{hubblerad}) and
(\ref{hubbletemp}) respectively. Correspondingly, the heat
capacity of the holographic statistics on outer screen can be
written as
\begin{eqnarray}
\label{CVb} C_V^b=c_v\frac{r_b^2}{G},
\end{eqnarray}
where $c_v$ is a constant of the order of ${\cal{O}}(1)$
determined by the detailed microscopic quantities of quantum
gravity.

Having expressed the heat capacity, we repeat the steps of the
previous subsection, that is we insert (\ref{CVb}) in
(\ref{deltarho2}) and then in (\ref{PPhi}), with all quantities
considered with a subscript $b$. We mention here that in this case
the correlation length coincides with the holographic screen's
radius $r_b$, as it has been shown from black-hole physics and its
application to cosmology \cite{Nayeri:2005ck, Magueijo:2006fu, Cai:2009rd}. Assembling everything we extract the
expression of the power spectrum for perturbations caused by
holographic fluctuations on the outer screen,namely
\begin{eqnarray}
P_{\Phi}^b=\frac{2c_v\beta^6}{\pi^2}GH^2. \label{PPhiH0}
\end{eqnarray}
Furthermore, in the specific model with $\beta=\sqrt{2}$, the
power spectrum writes as
\begin{eqnarray}\label{PPhiH}
P_\Phi^b=\frac{16c_v}{\pi^2}GH^2.
\end{eqnarray}
As we observe, the primordial curvature perturbations are also
scale-invariant.

Finally, we mention that one should repeat these calculations for
the inner screen, too. From ({\ref{PPhiH0}}) it is implied that
the power spectrum of holographic fluctuations is almost
proportional to one over the area of the screen, and thus the
contribution of the inner screen might be significant. However,
since the size of the inner screen is much smaller than the
cosmological scale during inflation, its corresponding
fluctuations only contribute to the sub-Hubble regime. Therefore,
we can neglect the inner screen contribution to CMB observations.

\subsection{Primordial perturbation spectrum}

In the previous two subsections we extracted the expressions for
the power spectrum for primordial curvature perturbations,
generated by the radiation sector, as well as by the outer
holographic screen, namely relations (\ref{radspect1}) and
(\ref{PPhiH0}) respectively, neglecting possible interaction terms
between the two perturbation sources and the inner screen
contribution. Comparing the two results we can immediately find
that the perturbation amplitude from radiation behaves like the
square of the one from the outer holographic screen. Fitting with
current CMB observations, one concludes that $P_{\Phi}^b$ is of
the order of ${\cal{O}}(10^{-10})$, and therefore $P_{\Phi}^r$ is
completely negligible. Thus, this is a significant difference of
the model at hand from conventional cosmology, that is the main
source of perturbation comes from the outer holographic screen and
not from the radiation sector of the universe interior. This
feature has some interesting physical implications.

Let us specify the discussion, considering a variable that is
widely used, namely the curvature perturbation in comoving
coordinates  \cite{Mukhanov:1990me}:
\begin{eqnarray}
\zeta\equiv\Phi+\frac{{\cal H}}{{\cal H}^2-{\cal H}'}(\Phi'+{\cal
H}\Phi)~.
\end{eqnarray}
This variable can be computed from the gravitational potential
$\Phi$ and background parameters. Since in inflation the metric
perturbation is frozen at super-Hubble scales, we acquire the
simple relation $\zeta\simeq\Phi/\epsilon$. As a consequence, and
using ({\ref{PPhiH0}}), we obtain the primordial power spectrum of
curvature perturbation during the thermally induced inflationary
period in entropic cosmology:
\begin{eqnarray}
 P_{\zeta}\simeq\frac{16c_v}{\epsilon^2\pi^2}GH^2~.
\end{eqnarray}
Therefore, we can easily calculate the spectral index, which is
the basic quantity in any relevant discussion
 \cite{Sasaki:1995aw}, namely
\begin{eqnarray}
\label{spectrond}
 n_S-1\equiv\frac{d\ln{P_{\zeta}}}{d\ln{k}}=-2\epsilon-2\eta~,
\end{eqnarray}
where $\eta\equiv\dot\epsilon/H\epsilon$. In the deviation of this
relation we have used the usual relation $d \ln k\approx
d\ln(aH)\approx d\ln(a)$ \cite{Mukhanov:1990me}, which results
from the fact that during inflation the variation of the scale
factor is much larger than that of the Hubble parameter.

\begin{figure}[htbp]
\includegraphics[scale=0.83]{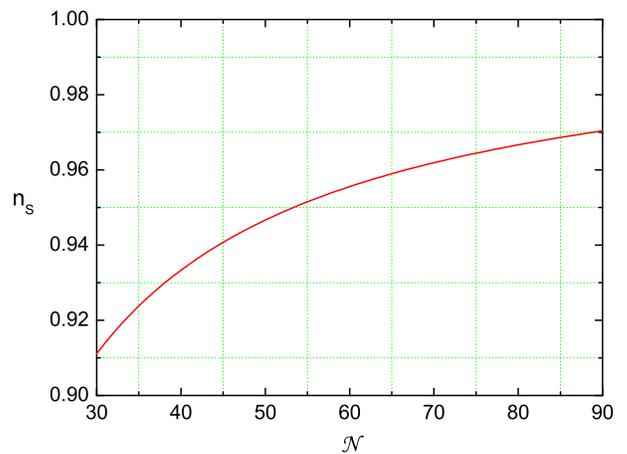}
\caption{{\it{The spectral index $n_S$ of primordial curvature
perturbation in the entropic scenario of thermal inflation as a
function of the efolding number ${\cal N}$. }}} \label{Fig:nS-N}
\end{figure}

Interestingly, in the scenario at hand one can approximately
extract a very simple relation connecting the spectral index of
thermal inflation $n_S$ with the efolding number ${{\cal N}}$. In
particular, combining (\ref{epsilondef}), (\ref{efolds}) and
(\ref{spectrond}), using $t_{i}$ as a free variable and using the
background parameter value $\beta=\sqrt{2}$, we result to
\begin{eqnarray}
\label{nSN}
 n_S\simeq1-\frac{8}{3{\cal N}}~.
\end{eqnarray}
Thus, it is obvious that the longer time inflation lasts, the
closer to scale-invariance will be the spectral index. Therefore,
we can immediately construct the $n_S$-${{\cal N}}$ graph, which
is presented in Fig.~\ref{Fig:nS-N}. Indeed, at large ${{\cal N}}$
the resulting spectrum is very close to scale-invariance, with a
red tilt, and the deviation from 1 is quantitatively in agreement
with observations, which require that $n_S = 0.96\pm0.012$ at
2$\sigma$ level  \cite{Komatsu:2010fb}.
This is a basic result of the present work, revealing that the
dominance of holographic fluctuations not only does not affect the
scale-invariant, conventional thermal ones, but it also improves
the picture of the produced spectrum.

Finally, let us examine the tensor perturbations and their
relation to the scalar ones examined above. Such a quantity, that
is the tensor-to-scalar ratio, is the second measure, along with
the spectral index, that characterizes the primordial
fluctuations. In the scenario of thermal inflation in entropic
cosmology, the primordial power spectrum for tensor perturbation
coincides with that of the usual slow-roll inflation, which reads
$P_T=16GH^2/\pi$  \cite{Sasaki:1995aw}, since the holographic
screens do not affect the tensor part of perturbation equations.
Therefore, defining as usual the ratio $r$ of tensor-to-scalar
perturbation, we acquire
\begin{eqnarray}
\label{te-to-sc}
 r\equiv\frac{P_T}{P_\zeta}=\frac{\epsilon^2\pi}{c_v}~,
\end{eqnarray}
which is doubly suppressed by the slow roll parameter $\epsilon$
but may be enhanced by a small value of the holographic parameter
$c_v$. This behavior is different from the usual inflationary
scenario. In particular, we conclude that in the general case with
$c_v\sim O(1)$ the primordial tensor perturbation is insensitive
to current cosmological observations, but it is still possible to
obtain sizable tensor modes if we fine-tune the value of $c_v$ to
be small enough.

\begin{figure}[htbp]
\includegraphics[scale=0.83]{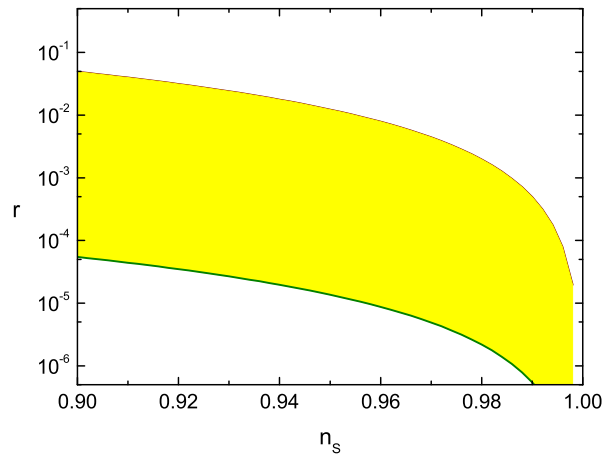}
\caption{{\it{The contour of the tensor-to-scalar ratio $r$ in the
entropic scenario of thermal inflation, as a function of the
spectral index $n_S$, in the value regime of $0.01<c_v<9$. }}}
\label{Fig:r-nS}
\end{figure}

To investigate these features in more detail we again combine
(\ref{epsilondef}) and (\ref{efolds}) using $t_{i}$ as a free
variable and using the parameter value $\beta=\sqrt{2}$, resulting
to the helpful relation $\epsilon\approx \frac{1}{3\cal N}$. Thus,
insertion into (\ref{te-to-sc}) leads to the simple relation
\begin{eqnarray}
\label{rN}
 r\approx \frac{\pi}{9 c_v {\cal N}^2}.
\end{eqnarray}
\begin{figure}[htbp]
\includegraphics[scale=0.83]{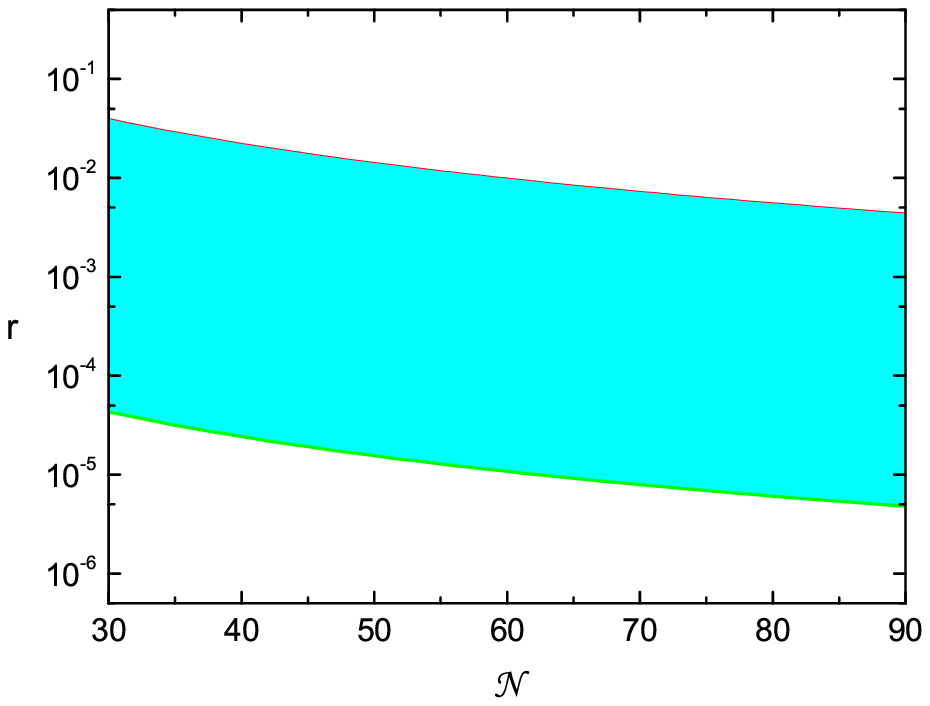}
\caption{{\it{The contour of the tensor-to-scalar ratio $r$ in the
entropic scenario of thermal inflation, as a function of the
efolding number ${\cal N}$, in the value regime of $0.01\leq
c_v\leq9$. }}} \label{Fig:r-N}
\end{figure}
In Fig. \ref{Fig:r-N} we present the  $r$-${{\cal N}}$ graph
taking the value regime of $c_v$ between $0.01$ and $9$. As we
observe, at large efolding ${{\cal N}}$ the resulting
tensor-to-scalar ratio $r$ acquires very small values.
Furthermore, combining (\ref{nSN}) with (\ref{rN}), we find
\begin{eqnarray}
\label{rnS}
 r\approx \frac{\pi(1-n_S)^2}{64 c_v}.
\end{eqnarray}
The corresponding $r$-$n_S$ graph is presented in Fig.
\ref{Fig:r-nS}, taking $c_v$ in the interval from $0.01$ to $9$.
Comparing this figure with latest cosmological data
\cite{Komatsu:2010fb, Xia:2008ex}, it is clear that
our results are compatible with current observations. Moreover,
the smallness of $r$ seems to be closer to observations comparing
to the usual paradigm of chaotic inflationary models
\cite{Linde:1983gd}.

\section{Non-Gaussianities}
\label{non-gauss}

Recently, a lot of interest has been paid on the analysis of
non-linear perturbations at early universe, under the scenarios of
single field slow-roll inflation  \cite{Maldacena:2002vr}, brane
inflation models  \cite{Alishahiha:2004eh, Lidsey:2007gq,
Cai:2009hw}, inflation models with general non-canonical form
\cite{Seery:2005wm,Arroja:2008ga}, curvaton configurations
\cite{Lyth:2001nq, Li:2008fma, Huang:2008ze}, Ekpyrotic scenarios
\cite{Koyama:2007if}, phantom inflation \cite{Piao:2004tq}, matter
bounce cosmology \cite{Cai:2009fn} etc (see \cite{Bartolo:2004if,
Chen:2010xk} for recent reviews). Such a non-linear analysis is
necessary in order to reveal the possible non-Gaussianity of the
primordial fluctuations, which can be measured by cosmological
observations \cite{Yadav:2007yy}. Thus, along with the examination
of the spectral index and the tensor-to-scalar ratio, the
estimation of the non-Gaussianities that are produced by an
inflationary scenario is a crucial step, since they can constrain
or rule out the examined scenario.

In this section we investigate the non-linear perturbation of
thermal inflation in double-screen entropic cosmology, by
computing its non-Gaussianity estimator. This technique of
incorporating non-Gaussianities in a thermal system has been
applied in an inflationary model coupled to normal radiation
\cite{Chen:2007gd}, in the context of bounce cosmology
\cite{Cai:2009rd}, in a string gas scenario \cite{Chen:2007js},
and in a holographic universe \cite{Ling:2008xd}.

For a perturbation mode with fixed $k$ its non-Gaussianity
estimator is given by the amplitude of the three-point correlation
function over the square of the two-point one, and can be written
as
\begin{eqnarray}\label{fnl}
 f_{NL}=\frac{5\langle\zeta_k^3\rangle}{18k^{3/2}\langle\zeta_k^2\rangle^2}~.
\end{eqnarray}
In the previous section we calculated the result of the two-point
function, namely relation (\ref{deltarho2}). Therefore, we need to
calculate also the three-point correlator.

As we showed above, the dominant contribution of primordial
curvature perturbation comes from the holographic fluctuations on
the outer boundary surface. Thus, the key point is to calculate
the three-point function of these holographic fluctuations. In an
equilibrium ensemble, one obtains  \cite{Cai:2009rd}
\begin{eqnarray}
\label{3pt}
 \langle\delta\rho_b^3\rangle|_{t_*}
 =\frac{T_b^2}{R_b^9}\frac{\partial}{\partial{T_b}}(C_V^bT_b^2)
 =\frac{2c_vT_b^3}{GR_b^7}~,
\end{eqnarray}
which is calculated at the moment $t_*$ of Hubble crossing in
coordinate space. In the above relation we have used  (\ref{CVb}),
and the fact that the correlation length $R_b$ coincides with the
holographic screen's radius $r_b$.

In summary, we can now insert (\ref{3pt}) and (\ref{deltarho2})
(applied with indices ``b'') into the expression of
non-Gaussianity estimator (\ref{fnl}). Using also the
approximation $\zeta\simeq\Phi/\epsilon$, we finally obtain
\begin{eqnarray}
\label{fNL1}
 f_{NL}\simeq\frac{5}{36\sqrt{2}\pi^2}\frac{\epsilon R_bH^2}{c_vT_b}~.
\end{eqnarray}
We mention here that this result is similar to the one obtained in
a thermal bouncing universe filled with Gibbons-Hawking radiation
\cite{Cai:2009rd}. However, there exists a manifest difference
between the two results, that is the non-Gaussianities may be
suppressed by the slow-roll parameter $\epsilon$ in the present
scenario of entropic thermal inflation, but not in the thermal
bouncing cosmology. Thus, we deduce that this suppression behavior
is a consequence and a physical reflection of inflation.

Proceeding forward, we insert in (\ref{fNL1}) the expression for
$R_b$, that is for $r_b$, which is given by (\ref{hubblerad}), and
for $T_b$, which is given by (\ref{hubbletemp}), resulting to
\begin{eqnarray}
 f_{NL}\simeq\frac{5\epsilon}{36\sqrt{2}\pi c_v}~.
\end{eqnarray}
This relation provides the non-Gaussianity of thermal inflation in
double-screen entropic cosmology, and as we observe it is
scale-invariant.

We close this section by mentioning that although the
non-Gaussianities of primordial curvature perturbation in the
scenario at hand are suppressed by the slow-roll parameter
$\epsilon$, it is still possible to produce a sizable value of
$f_{NL}$ if $c_v\sim\epsilon$. Therefore, in order to complete a
quantitative investigation, one needs to perform a detailed
analysis of the microscopic properties of a holographic screen in
entropic cosmology, which will lead to an estimation of $c_v$.
However, such an analysis is a hard task under the present
knowledge, and it lies outside the scope of the present work.

\section{Conclusion}
\label{Conclusion}

In this work we investigated a scenario of thermal inflation
realized by two holographic screens in the context of entropic
cosmology. We found that the realization of inflation is general,
resulting from the system evolution from non-equilibrium to
equilibrium. Going beyond the background evolution, we analyzed
the primordial curvature perturbations arising from the universe
interior, as well as the thermal fluctuations generated on the
outer holographic screen. For both these contributions the power
spectra are almost scale-invariant, however the amplitude of
curvature perturbation arisen from holographic fluctuations on the
outer screen is much larger than that of the universe interior.
Furthermore, due to the thermal initial conditions for scalar-type
metric perturbations, the consistency relation widely held in
usual inflation models was found to be modified in the present
scenario. In summary, the produced power spectrum is nearly
scale-invariant with a red tilt.

Proceeding forward, we provided approximate analytic expressions
for the tensor-to-scalar ratio as a function of the spectral
index, with the one free parameter $c_v$ determined by the
detailed microscopic quantities of quantum gravity. As we saw, the
corresponding contour plot is in agreement with observations, with
even better quantitative features comparing to the usual paradigm
of chaotic inflationary models.

Finally, we examined the non-Gaussian distribution of the
inhomogeneities of primordial curvature perturbations, generated
from the outer screen. Since these fluctuations satisfy the
holographic statistics, the resulting non-linearity parameter is
inversely proportional to $c_v$, and it is suppressed by the
slow-roll parameter, while it is nearly scale-invariant.
Therefore, a sizable value of the non-linearity parameter is
possible due to holographic statistics on the outer screen,
provided $c_v$ is of the same order with the slow-roll parameter.

It is important to mention that our analysis involves a few
uncertainties on the coefficients, since the detailed
thermodynamics of holographic statistics on the boundary screens
is still not well understood in current knowledge. This provides a
wide parameter space to fit to current cosmological observations.
Therefore, it may be far from conclusive to give strong
constraints on the scenario of double-screen entropic thermal
inflation. However, we still might be able to distinguish such a
scenario from a normal model of slow-roll inflation by measuring
the spectral indexes of primordial power spectra and examining
their consistency relation in the coming experiments. Moreover, we
expect that the scenario considered in this work can be
theoretically developed along with the accumulating studies on
holographic properties of entropic cosmology, so that it may be
verified or ruled out by future cosmological data.

As an end, we would like to point out that a distinguishable
feature of entropic cosmology with double holographic screens is
the explanation of inflation and late time acceleration in a
unified frame. In this work we focused on the predictions of
inflation realized by holographic screens out of thermal
equilibrium at early universe. However, at significantly late
times the inner screen would evaporate and thus yield another
acceleration epoch, which could explain the current dark-energy
period. Therefore, we expect that the scenario at hand might be
related to the holographic dark energy scenario, which
incorporates the universe acceleration in consistency with the
basic quantum gravitational requirements embedded in the
holographic principle \cite{HDE}.

\section*{Acknowledgments}

It is a pleasure to thank D. A. Easson, H. Li, T. Qiu, Y. Wang, and
T. Vachaspati for helpful discussions and comments. YFC thanks
the Institute for the Physics and Mathematics of the Universe and
the Research Center for the Early Universe at the University of
Tokyo, Tokyo University of Science, and the Yukawa Institute for
Theoretical Physics at Kyoto University for their hospitality when
this work was initiated. The research of YFC is supported in part by
the Arizona State University Cosmology Initiative. ENS thanks the
Physics Department of National Tsing Hua University of Taiwan, for
the hospitality during the preparation of this work.



\end{document}